\documentclass{aa}
\usepackage{graphicx}
\usepackage{txfonts}
\begin{document}
   \title{Gaussian decomposition of \ion{H}{i} surveys}
   \subtitle{III. Local \ion{H}{i}}
   \author{U. Haud
           \inst{1}
        \and
           P. M. W. Kalberla
           \inst{2}
           }

   \offprints{U. Haud}

   \institute{Tartu Observatory, 61\,602 T\~oravere, Tartumaa, Estonia\\
              \email{urmas@aai.ee}
           \and
              Argelander-Institut f\"ur Astronomie, Universit\"at Bonn
              \thanks{Founded by merging of the Sternwarte,
              Radioastronomisches Institut and Institut f\"ur
              Astrophysik und Extraterrestrische Forschung der
              Universit\"at Bonn}, Auf dem H\"ugel 71, 53\,121 Bonn,
              Germany\\
              \email{pkalberla@astro.uni-bonn.de}
              }
   \date{Received \today; accepted \today}
   \abstract
      {To investigate the properties of the 21-cm radio-lines of
      galactic neutral hydrogen, the profiles of ``The
      Leiden/Argentine/Bonn (LAB) Survey of Galactic \ion{H}{i}'' are
      decomposed into Gaussian components.}
      {The width distribution of the obtained components is analysed
      and compared with similar studies by other authors.}
      {The study is based on an automatic profile decomposition
      algorithm. As the Gaussians obtained for the complex \ion{H}{i}
      profiles near the galactic plane cannot be directly interpreted
      in terms of the properties of gas clouds, we mainly study the
      selected simplest profiles in a limited velocity range. The
      selection criteria are described and their influence on the
      results is discussed.}
      {Considering only the simplest \ion{H}{i} profiles, we
      demonstrate that for Gaussians with relatively small LSR
      velocities ($-9 \le V_\mathrm{C} \le 4~\mathrm{km\,s}^{-1}$) it
      is possible to distinguish three or four groups of preferred
      line-widths. The mean widths of these groups are $\mathrm{FWHM} =
      3.9 \pm 0.6$, $11.8 \pm 0.5$, $24.1 \pm 0.6$, and $42 \pm
      5~\mathrm{km\,s}^{-1}$. Verschuur previously proposed similar
      line-width regimes, but with somewhat larger widths. He used a
      human-assisted decomposition for a nearly 50 times smaller
      database and we discuss systematic differences in analysis and
      results.}
      {The line-widths of about 3.9 and $24.1~\mathrm{km\,s}^{-1}$ are
      well understood in the framework of traditional models of the
      two-phase interstellar medium. The components with the widths
      around $11.8~\mathrm{km\,s}^{-1}$ indicate that a considerable
      fraction (up to about 40\%) of the \ion{H}{i} gas is thermally
      unstable. The reality and the origin of the broad lines with the
      widths of about $42~\mathrm{km\,s}^{-1}$ is more obscure.
      These, however, contain only about 4\% of the total observed
      column densities.}

      \keywords{ISM: atoms~-- solar neighbourhood~-- Radio lines: ISM}

   \maketitle

   \section{Introduction}

      The tradition to analyse the 21-cm \ion{H}{i} line profiles in
      terms of Gaussian components has its roots in the finding that
      the \ion{H}{i} absorption features are often well represented by
      Gaussian functions in optical depth $\tau(\nu)$. This is
      understandable when the line profile is dominated by the thermal
      motions or nonthermal motions consisting of a large number of
      turbulent elements. The decomposition of the emission profiles is
      physically meaningful only when $\tau(\nu) \ll 1$. Nevertheless,
      both the opacity and brightness profiles of most sources are
      easily decomposed into Gaussains, so whether or not this model is
      physically correct, it works empirically and is convenient.
      Moreover, even if such a decomposition is not a physically valid
      description of the galactic ISM, it is a valid mathematical
      description of the corresponding 21-cm \ion{H}{i} profile.

      Mathematically the most questionable point of the Gaussian
      decomposition is that often it is not unique. This nonuniqueness
      means that several quite different solutions may approximate the
      observed profile almost equally well, and the decomposition
      provides no satisfactory means for choosing between these
      solutions, while others, equally good or even better ones, may
      not be found at all. However, when we decompose a large number of
      profiles by exactly the same algorithm, we may hope that for
      similar profiles the algorithm will prefer similar
      decompositions. This means that usually some specific features in
      the observed profiles are represented by some specific set of
      Gaussians, which can be found from the overall data-set more
      easily than un-parametrized spectral features. The latter is one
      of our main interests in Gaussian decomposition and from this
      point of view all obtained Gaussians are real, as they describe
      some features in observed profiles. If and how these Gaussians
      are related to the physical properties of the galactic ISM is a
      question, which must be answered separately for every selected
      subset of the components.

      Proceeding from these considerations, in the first paper of this
      series (Haud \cite{Hau00}, hereafter Paper~I) we described a new
      fully automated computer programme for the decomposition of large
      21-cm \ion{H}{i} line surveys into Gaussian components. We used
      the Leiden/Dwingeloo Survey (LDS) of galactic neutral hydrogen
      (Hartmann \cite{Har94}) as test data for decomposition. After our
      initial paper a new revision of the LDS data (LDS2) has been
      published (Kalberla et al. \cite{Kal05}). Recently a similar
      Southern sky high sensitivity \ion{H}{i} survey at $\delta \le
      -25\degr$ was completed in the Instituto Argentino de
      Radioastronomia (IARS) and published by Bajaja et al.
      (\cite{Baj05}). Both surveys have been combined to form the
      Leiden/Argentine/Bonn (LAB) full sky database (Kalberla et al
      \cite{Kal05}). Using our Gaussian decomposition programme, we
      have decomposed all profiles in these surveys.

      The specifications for the LDS2 and the IARS closely match each
      other, but all the data reduction and calibration procedures for
      combining them into the LAB were carried out entirely
      independently (Kalberla et al. \cite{Kal05}). Proceeding from
      this, also the Gaussian decomposition was carried out separately
      for the LDS2 and the IARS. For the published version of the LDS2,
      in the cases of repeated observations at the same sky positions,
      the final profiles were selected on the basis of the best
      agreement of their preliminary Gaussian decompositions with the
      decompositions of the neighbouring profiles (Kalberla et al.
      \cite{Kal05}) and for our final decomposition we used these
      preselected profiles. In the case of the IARS, we used for the
      decomposition the original 1008-channel data of all the observed
      profiles and for repeated observations we applied before the
      final decomposition the same selection criteria as used for the
      LDS2.

      In both surveys the observations were made on a galactic
      coordinate grid with observed points spaced by $(\Delta l, \Delta
      b) = (0 \fdg 5/\cos b, 0 \fdg 5)$. The velocity resolution was
      $1.03~\mathrm{km\,s}^{-1}$ in the case of the LDS2 and
      $1.27~\mathrm{km\,s}^{-1}$ for the IARS. For both surveys the
      authors declare the final rms noise of the data to be about
      $0.07~\mathrm{K}$, but our estimates have given somewhat larger
      values. For the LDS we have obtained an estimate of
      $0.09~\mathrm{K}$ (Paper~I) and the same value is given also by
      Westphalen (\cite{Wes97}). With the IARS the situation seems to
      be more complicated (Haud \& Kalberla \cite{Hau06}, hereafter
      Paper~II). When for the signal-free regions of the profiles the
      estimate given by the authors of the survey seems to be correct,
      in the regions containing the emission signal, we estimated
      higher noise levels than it may be expected on the basis of the
      signal-free regions and the radiometer equation. From the regions
      containing the line emission we estimated the noise level for
      zero intensity line emission to be about $0.08~\mathrm{K}$
      (Fig.~12. of Paper~II).

      When performing the Gaussian decomposition, the main attention
      was turned on keeping the number of resulting Gaussians as small
      as possible (Paper~I). For every profile in the first
      decomposition stage new Gaussians were added until the rms of
      the residuals became less or equal to the noise level of the
      signal-free regions of the profile. The dependence of the noise
      level on the signal strength was taken into account according to
      the radiometer equation. After preliminary decomposition of each
      profile special analysis was applied to find additional
      possibilities for removing some less important components from
      the final decomposition without increasing the residuals above
      the noise level. To reduce the ambiguities in choosing between
      different, but nearly equally acceptable solutions, we have used
      the assumption that in survey observations the profiles from
      neighbouring sky positions must share some common properties and
      therefore their decompositions must be as similar as possible.
      The described procedure gave us 1\,064\,808 Gaussians per
      138\,830 profiles for the LDS2 and 444\,573 Gaussians per 50\,980
      profiles in the case of the IARS.

      In Paper~II we analysed the distributions of the parameters of
      the obtained components. We discussed both the original LDS data
      and the newer LAB. The main attention was paid to the separation
      of the Gaussians describing different artefacts of the
      observations (interferences), reduction (baseline problems) and
      the decomposition (separation of the signal from the noise)
      process. In the present paper, we continue the analysis of the
      distribution of the Gaussian parameters, but in this case the
      main focus is on the components, most likely corresponding to the
      galactic \ion{H}{i}. We start with the general distribution of
      all Gaussians and concentrate on the distribution of the widths
      of weak components with near zero velocities.

      Earlier, a similar distribution has extensively been studied by
      Verschuur and co-authors (\cite{Ver89}, \cite{Ver94},
      \cite{Ver99}, \cite{Ver02}, \cite{Ver04}). Their results are
      qualitatively very similar to ours, but we must note some
      differences in the quantitative values. By trying to follow the
      approach used by Verschuur, we point out the main differences
      found and their possible influence on the final results. In the
      concluding sections of the paper, we discuss the relation of our
      results to the properties of the galactic \ion{H}{i}.

   \section{The general distribution of Gaussian widths}

\begin{figure}
         \resizebox{\hsize}{!}{\includegraphics{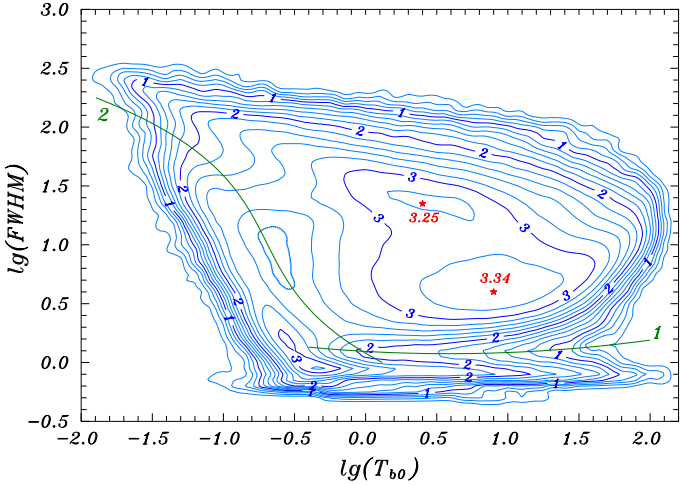}}
         \caption{Frequency distribution of the parameter values in
            $(\lg(T_\mathrm{b0}), \lg(\mathrm{FWHM}))$ plane for all
            Gaussians obtained in the decomposition of the LAB.
            Isodensity lines are drawn in the scale of $\lg(N+1)$ with
            the interval of 0.2. The two main maxima are labelled with
            their height values. The green thick solid lines 1 and 2
            represent the selection criteria discussed in Paper~II.}
         \label{Fig01}
      \end{figure}

      In Paper~II we discussed mainly the frequency distribution of the
      Gaussians in their height vs. width plane (Fig.~\ref{Fig01}). We
      defined the height of a Gaussian by the value of the central
      brightness temperature $T_\mathrm{b0} > 0$ from the standard
      Gaussian formula
      \begin{equation}
         T_\mathrm{b} = T_\mathrm{b0}
                   \mathrm{e}^{-\frac{(V - V_\mathrm{C})^2}
                   {2\sigma_V^2}}, \label{Eq1}
      \end{equation}
      where $T_\mathrm{b}$ is the brightness temperature and $V$ is the
      velocity of the gas relative to the Local Standard of Rest.
      $V_\mathrm{C}$ is the velocity corresponding to the centre of the
      Gaussian. We characterise the widths of the components by their
      full width at the level of half maximum ($\mathrm{FWHM}$), which
      is related to the velocity dispersion $\sigma_V$ obtained from
      our decomposition programme by a simple scaling relation
      $\mathrm{FWHM} = \sqrt{8 \ln{2}} \sigma_V$.

      In Paper~II we have demonstrated that not all Gaussians obtained
      in our decomposition and represented in Fig.~\ref{Fig01} could be
      considered as corresponding to the real \ion{H}{i} emission. A
      considerable number of the obtained components are due to
      different observational, reductional and decompositional
      problems. Very narrow Gaussians actually represent the strongest
      random noise peaks misinterpreted by the decomposition programme
      as possible signal peaks and the radio-interferences not found or
      properly removed during the reduction of the observed profiles.
      Many of somewhat wider weak Gaussians seem to be caused by the
      increased uncertainties in bandpass removal near the profile
      edges. In the direction of even wider Gaussians, at least some
      weak components may be due to the problems in the determination
      of the profile baselines, but some of these Gaussians may also
      arise from the emission of the high velocity dispersion halo gas
      (HVDHG) reported by Kalberla et al. (\cite{Kal98}).

      In Fig.~\ref{Fig01} such spurious components (in the lower
      left-hand part of the figure) are separated by two thick solid
      lines from the others (in the upper right-hand part of the
      figure) more likely describing the real galactic \ion{H}{i}
      emission. However, from Fig.~\ref{Fig01} we can also see that the
      parameters of the components, most likely corresponding to the
      galactic \ion{H}{i}, are not distributed randomly, but exhibit
      concentration around two more or less distinct values. We can
      clearly see such a concentration around the values of
      $T_\mathrm{b0} \approx 8~\mathrm{K}$, $\mathrm{FWHM} \approx
      4~\mathrm{km\,s}^{-1}$ and a weaker one at $T_\mathrm{b0} \approx
      2.5~\mathrm{K}$, $\mathrm{FWHM} \approx 22~\mathrm{km\,s}^{-1}$.
      This finding is in general agreement with the earlier results by
      Mebold (\cite{Meb72}), who decomposed nearly 1700 21-cm
      \ion{H}{i} emission line profiles into about 2400 Gaussian
      components and found that their width distribution has two
      distinct maxima at $\mathrm{FWHM} \approx 7~\mathrm{km\,s}^{-1}$,
      mostly populated by components with $10 \la T_\mathrm{b0} \la
      50~\mathrm{K}$ (narrow components), and $24~\mathrm{km\,s}^{-1}$
      for Gaussians with $3 \la T_\mathrm{b0} \la 10~\mathrm{K}$
      (shallow components).

      Mebold (\cite{Meb72}) used the profiles with lower spectral
      resolution than that of the LAB. Moreover, his profiles were from
      relatively low galactic latitudes and the stronger emission
      components become wider and more non-Gaussian due to the
      saturation and self-absorption. Considering this, we have a
      surprisingly good match of the maxima of at first glance
      physically meaningless (lumping into one high-latitude \ion{H}{i}
      and heavily-saturated low-latitude galactic disk profiles, HVC's
      and IVC's and even the observational, reductional and
      decompositional artefacts) distribution in Fig.~\ref{Fig01} with
      the ones described by Mebold. In his paper Mebold concluded:
      ``... it is most likely that the narrow components are emitted by
      cold ($T \la 70~\mathrm{K}$) \ion{H}{i} clouds and the shallow
      components by a hot ($750 \la T \la 9200~\mathrm{K}$) and
      optically thin \ion{H}{i} intercloud medium''. However, in this
      way the general coincidence of the distribution properties of the
      Gaussians described above, may indicate that even in our
      mathematical representation of the profiles with the Gaussians at
      least some components may preserve physical information about the
      properties of the underlying galactic \ion{H}{i}. In the
      following we will try to separate from the overall set of
      Gaussians the ones most directly related to the structure of the
      local galactic \ion{H}{i}.

   \section{The central velocity vs. width distribution}

      It is well known that due to the concentration of the gas in the
      Galaxy into a thin disk and the differential rotation of this
      disk, the parameters of the observed \ion{H}{i} profiles depend
      rather strongly on the direction of the observations. The area
      under the profile shows the $1 / \sin b$ variation with the
      galactic latitude. The centre of gravity of the line has the
      $\sin 2l$ variation with longitude and $\cos^2 b/\sin b$
      variation with latitude. The line-width has a component which
      changes like $\sin^2 2l$ with longitude and like $\cos^4 b/\sin^2
      b$ with latitude. A considerable part of these changes in
      line-shapes are represented in the Gaussian decomposition by
      decomposing the more complicated profiles near the galactic plane
      with the larger numbers of Gaussians, but most likely to a
      certain extent these dependences will also affect the properties
      of the Gaussians themselves.

      \begin{figure*}
      \sidecaption
         \includegraphics[width=12cm]{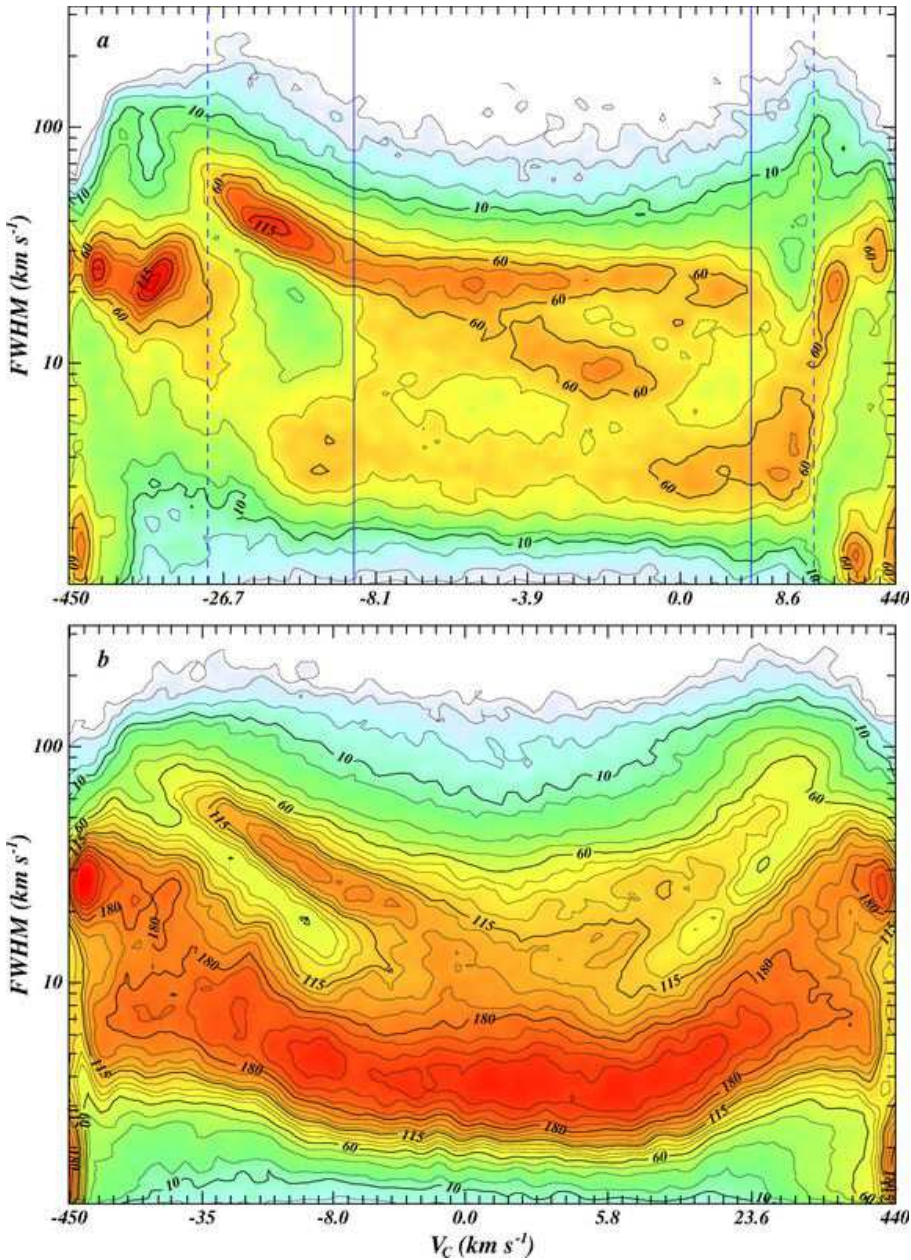}
         \caption{The distribution of Gaussian parameters in the
            central velocity -- Gaussian width plane for two different
            latitude ranges: upper (a) panel for $|b| \ge 40\degr$ and
            the lower (b) one for $|b| < 40\degr$. The contour lines on
            both panels are drawn at the levels of 1, 2, 5, 10, 15, 30,
            45, 60, 75, 85, 100, 115, 125, 140, 160, 180, 200, 225, 250
            and 300 Gaussians per counting bin. The ticks on the x-axis
            are drawn so that the number of Gaussians between every
            pair of neighbouring ticks is approximately the same. The
            precise values of the velocity corresponding to ticks of
            the upper panel are: $-450$, $-215$, $-120$, $-80$, $-60$,
            $-52$, $-47$, $-41$, $-36$, $-30.8$, $-26.7$, $-22.5$,
            $-19.1$, $-16.5$, $-14.4$, $-12.7$, $-11.3$, $-10.2$,
            $-9.4$, $-8.7$, $-8.1$, $-7.6$, $-7.1$, $-6.6$, $-6.2$,
            $-5.8$, $-5.4$, $-5.0$, $-4.6$, $-4.2$, $-3.9$, $-3.5$,
            $-3.16$, $-2.82$, $-2.47$, $-2.13$, $-1.77$, $-1.39$,
            $-0.95$, $-0.50$, 0.00, 0.59, 1.33, 2.21, 3.2, 4.6, 6.3,
            8.6, 12.4, 21.6, 37, 70, 118, 207 and 440. Those for the
            lower panel are: $-450$, $-138$, $-107$, $-89$, $-74$,
            $-63$, $-54$, $-48$, $-41$, $-35$, $-30.3$, $-25.7$,
            $-22.0$, $-18.7$, $-15.8$, $-13.3$, $-11.3$, $-9.5$,
            $-8.0$, $-6.7$, $-5.5$, $-4.5$, $-3.6$, $-2.75$, $-1.97$,
            $-1.26$, $-0.59$, 0.00, 0.54, 1.08, 1.60, 2.10, 2.63, 3.2,
            3.8, 4.4, 5.1, 5.8, 6.6, 7.5, 8.5, 9.7, 11.1, 12.6, 14.5,
            16.9, 19.9, 23.6, 28.5, 35, 43, 52, 62, 73, 88, 107, 156
            and 440. In panel a, the vertical solid lines enclose the
            velocity region $-9 \le V_\mathrm{C} \le
            4~\mathrm{km\,s}^{-1}$ of weak dependence of the
            line-widths on the central velocity of the components and
            the dashed lines correspond to the region $-31 \le
            V_\mathrm{C} \le 18~\mathrm{km\,s}^{-1}$ discussed in the
            latter half of the paper.}
         \label{Fig04}
      \end{figure*}
      From the relations discussed above, it follows that with the
      increasing mean velocity of the line also the line-width must
      increase. For our Gaussians this is illustrated in
      Fig.~\ref{Fig04} separately for profiles at latitudes $|b| \ge
      40\degr$ (upper panel) and $|b| < 40\degr$ (lower panel). To
      construct both panels of this figure, we sorted in corresponding
      latitude intervals all Gaussians with central velocities in the
      decomposition range ($-460 < V_\mathrm{C} <
      396~\mathrm{km\,s}^{-1}$ in the LDS2 and $-437 < V_\mathrm{C} <
      451~\mathrm{km\,s}^{-1}$ in the IARS; for details see Papers~I
      and II) in the ascending order of their central velocities and
      then grouped the sequence into 129 bins of an equal number of
      Gaussians, rejecting some Gaussians with the most extreme
      velocities. We binned the line-widths in equal steps of
      $\lg(\sigma_V)$ of 0.025. The isolines in Fig.~\ref{Fig04} give
      the numbers of Gaussians in each of such two-dimensional
      parameter interval. The latitude limit of $|b| = 40\degr$ was
      chosen rather arbitrarily. For testing purposes, we constructed
      similar plots also by taking the limit at $|b| = 30\degr$. In
      this case, the total number of Gaussians was increased in the
      upper panel and decreased in the lower panel, but the general
      structure of the distributions remained the same. Therefore, we
      may expect that the exact value of the limiting latitude is not
      very critical until it remains in some reasonable range.

      The first thing we can see from Fig.~\ref{Fig04} is that the
      typical line-widths of the density enhancements in the
      distribution of Gaussian parameters are not independent of the
      velocities of the Gaussians, but are larger for larger absolute
      values of the central velocities of the Gaussians. Moreover, this
      dependence seems to be stronger for lower latitudes than for
      higher ones, as expected from the relations discussed above. We
      can also see that at high velocities Figs.~\ref{Fig04}a ($|b| \ge
      40\degr$) and \ref{Fig04}b ($|b| < 40\degr$) are rather similar
      (when we take into account the differences in the velocity scales
      of these plots). This is most likely caused by the fact that even
      near the galactic plane the amounts of the very high velocity gas
      remain modest, the corresponding Gaussians are relatively well
      separated from the bulk of the disk gas and their parameters are
      rather well defined.

      At lower velocities, the differences are more remarkable. As
      discussed in detail in Paper~II, there is much more gas near the
      galactic plane than at greater heights and therefore at low
      latitudes the effects of velocity crowding, blending and
      saturation become more severe and the profiles become more
      complex than at higher latitudes. As a result, the lower panel of
      Fig.~\ref{Fig04} is dominated by the narrowest Gaussians and at
      larger widths we cannot distinguish other clear frequency
      enhancements. Most likely such a picture is caused by the fact
      that in this region the obtained Gaussians are only
      mathematically representing the profile and do not have any
      direct relation to the properties of the galactic \ion{H}{i}.
      However, in the distribution of Gaussian widths at high galactic
      latitudes we can easily distinguish three frequency enhancements.
      Those with the smallest and largest widths correspond more or
      less to the enhancements visible also in the general distribution
      given in Fig.~\ref{Fig01}, but the concentration at intermediate
      widths is new, not unambiguously visible in the previous figure.

      When the line-widths of the density enhancements in
      Figs.~\ref{Fig04} are in general dependent of the central
      velocity of the corresponding Gaussians, they are nearly velocity
      independent in the region of $-9 \la V_\mathrm{C} \la
      4~\mathrm{km\,s}^{-1}$ (marked in Fig.~\ref{Fig04}a by two solid
      vertical lines). Therefore, in this velocity interval it would be
      also meaningful to estimate the mean line-widths of the three
      visible line-width groups of Gaussians. To obtain these
      estimates, we first found the distribution of the widths of all
      Gaussians with $|b| \ge 40\degr$, $-9 \la V_\mathrm{C} \la
      4~\mathrm{km\,s}^{-1}$ and modelled this distribution as a sum of
      three lognormal distributions defined by
      \begin{equation}
         f_{\mu, \sigma^2} (x) = \left\{
         \begin{array}{ll}
            0 & (x \le 0)\\
            \frac{1}{\sqrt{2\pi}\sigma x}
            \mathrm{e}^{-\frac{(\ln(x) - \mu)^2}
            {2\sigma^2}} & (x > 0)\,.
         \end{array}\right. \label{Eq2}
      \end{equation}
      From many different density functions, more or less suitable for
      modelling the distribution of Gaussian widths, the values of
      which are defined to be positive, we have chosen to use the
      lognormal distribution, as its usage was technically rather
      convenient (if some parameter is distributed according to the
      lognormal law, the distribution of the logarithms of this
      parameter is described by a Gaussian). Mebold (\cite{Meb72}) has
      argued that his narrow and shallow components are more easily
      distinguishable when the line-width distribution is studied in
      logarithmic scale and in comparison with all initially tested
      distributions the lognormal one gave the best fits.

      For the lognormal distribution, the mean value can be calculated
      as $\mathrm{M}(x) = \mathrm{e}^{\mu + \frac{\sigma^2}{2}}$, where
      $\mu$ and $\sigma$ may be considered as free parameters for
      fitting the model distribution to the observed one. In the case
      of Fig.~\ref{Fig04}a, we obtain for the means of three fitted
      distributions $\mathrm{FWHM} = 3.89 \pm 0.02$, $11.3 \pm 0.1$ and
      $23.8 \pm 0.2~\mathrm{km\,s}^{-1}$ (the error estimates here are
      the formal errors of the fit and do not account for errors in
      determination of the parameters of Gaussians or any possible
      systematic errors due to the applied data reduction methods).

      The exact numerical values for the mean line-widths of the near
      zero velocity density enhancements in Fig.~\ref{Fig04}a also
      depend on the used velocity limits (solid vertical lines in
      Fig.~\ref{Fig04}a). These limits in their turn depend somewhat on
      the applied latitude limit for the separation of the panels in
      Fig.~\ref{Fig04}: the lower latitude limits tend to indicate more
      symmetrical velocity limits. However, for reasonable changes in
      latitude and velocity limits the changes in the width estimates
      are rather small. For example, when lowering the latitude limit
      to $|b| = 30\degr$ and using the velocity limit $|V_\mathrm{C}|
      \la 8~\mathrm{km\,s}^{-1}$, we obtained the following mean width
      estimates: $\mathrm{FWHM} = 3.78 \pm 0.03$, $10.8 \pm 0.2$ and
      $23.9 \pm 0.5~\mathrm{km\,s}^{-1}$. The comparison of these
      numbers with the results given above also yields more realistic
      estimates of the actual reliability of the mean widths than the
      pure formal fitting errors.

   \section{The Gaussians from the simplest profiles}

      In the previous sections we used all the Gaussians obtained in
      the decomposition of all the profiles in the LAB database and
      tried to demonstrate how the distributions of their widths differ
      in different velocity or latitude regions. Finally, we
      concentrated our attention to the components with nearly zero
      velocity and to the relatively simple profiles at high galactic
      latitudes. Now we try to use the obtained information and limit
      ourselves to the simplest profiles and to the Gaussians
      describing only the low-velocity (relatively nearby) gas.
      Therefore, we must apply some selection criteria similar to those
      discussed in the previous section. First of all, as demonstrated
      by Fig.~\ref{Fig04}, we must exclude the complex profiles near
      the galactic plane. However, due to different complexity of the
      profiles in different longitude intervals the usage of a fixed
      latitude limit is probably not the best solution for this
      problem. Therefore, we decided to estimate the complexity of each
      profile individually.

      \begin{figure}
         \resizebox{\hsize}{!}{\includegraphics{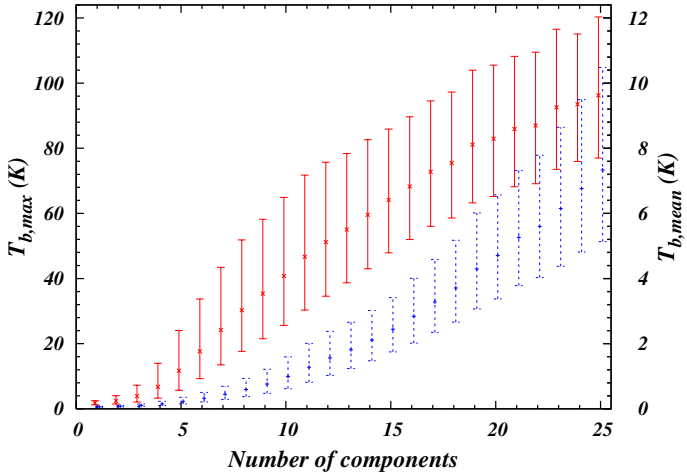}}
         \caption{The relation between the number of Gaussians used
            for the decomposition of the profile, the peak brightness
            temperature (as estimated from the Gaussian decomposition)
            of the profile (red solid symbols, the vertical axis on the
            left) and the mean channel value in the region used for
            decomposition (blue dashed symbols, the vertical axis on
            the right). The errorbars represent the standard deviations
            around the mean of the corresponding distributions. The
            symbols are horizontally shifted to avoid overlap.}
         \label{Fig05}
      \end{figure}
      The complexity of the profile has two aspects. On the one hand,
      it is determined by the line formation process and to consider
      this, we must know the properties of the gas along the line of
      sight. In general, this information is not available and
      therefore we cannot directly distinguish the lines representing
      the superposition of the emission of many gas concentrations
      projected on each other from those formed by the emission of a
      relatively small number of gas clouds. However, on average,
      the brighter profiles are also more complex.

      On the other hand, it is also important to know how well
      determined the parameters of the Gaussians obtained in the
      decomposition of every given profile are. In principle, this
      could be estimated. We have determined for every profile a set of
      Gaussian parameters, which corresponds to some minimal value of
      the residual rms of the deviations between the observed profile
      and its Gaussian representation. If we change the values of
      Gaussian parameters, the rms increases. The region within which
      the rms increases by no more than some predefined amount defines
      the confidence region around our solution. If we define the
      allowed increase of the rms so that the corresponding part of the
      parameter space contains some fixed percentage of the probability
      distribution of Gaussian parameters, the size and the shape of
      the confidence region give us the estimates for the errors of the
      obtained parameters. At the same time, this procedure requires
      huge amounts of computation, most likely not justified for the
      present study. However, on average, for profiles with a larger
      number of overlapping Gaussians the parameters of the latter are
      determined with lower confidence. Finally, as the number of
      Gaussians used for decomposition is rather well correlated with
      the emission content of the profile (Fig.~\ref{Fig05}), we choose
      as a primary indicator of the profile complexity the number of
      Gaussians needed for its decomposition.

      However, in Paper~II we demonstrated that some of the Gaussians
      are caused by different observational, reductional and
      decompositional problems and with high probability do not
      describe the properties of the galactic \ion{H}{i}. As some
      profiles contain many spurious Gaussians, their inclusion in
      the analysis may considerably distort the profile complexity
      estimates. Therefore, before considering the complexity of the
      profiles, it is useful to exclude such Gaussians from
      consideration. We identify them by the criteria illustrated in
      Fig.~\ref{Fig01} by thick solid lines. As described in Paper~II,
      these criteria are analytically given by
      \begin{equation}
         \lg(\mathrm{FWHM}) < 0.057\lg(T_\mathrm{b0})^2 -
                              0.067\lg(T_\mathrm{b0}) +
                              0.094 \label{Eq3}
      \end{equation}
      (very narrow Gaussians are mostly due to stronger observational
      noise peaks and radio-interferences) and
      \begin{equation}
         \begin{array}{ll}
         \lg(T_\mathrm{b0}) < & -0.370\lg(\mathrm{FWHM})^3 +
                                 1.132\lg(\mathrm{FWHM})^2 -\\
                              & 1.567\lg(\mathrm{FWHM}) +
                                 0.117\\
         \end{array}
         \label{Eq4}
      \end{equation}
      (most of the weakest components are caused by different
      uncertainties in the determination of the baselines for
      profiles). Here we are not interested in high-velocity clouds and
      therefore do not apply in the case of Eq. (\ref{Eq4}) any
      velocity limits used in Paper~II, but we still reject all
      components with the central velocities outside the velocity range
      used for the decomposition. After rejection of these spurious
      components we classified the profiles on the basis of the total
      number of remaining Gaussians needed for the decomposition of any
      particular profile.

      So far we have used in this section all the Gaussians
      irrespective of their central velocities. However, in the
      previous section we demonstrated that the mean line-widths of the
      components tend to increase with the increase of the absolute
      value of their central velocities. We also saw that this increase
      is slower for near zero velocities and becomes stronger at higher
      velocities. It is somewhat arbitrary to fix any numerical
      velocity limits for the regions with slow and faster line-width
      changes. In Fig.~\ref{Fig04} these differences are also slightly
      exaggerated by the compression of the velocity scale at higher
      velocities. Nevertheless, when for the line-width group with
      widest Gaussians in Fig.~\ref{Fig04}a in the velocity interval of
      $|V_\mathrm{C}| \la 5~\mathrm{km\,s}^{-1}$ the mean line-width
      changes are not more than $0.25~\mathrm{km\,s}^{-1}$ per
      $1~\mathrm{km\,s}^{-1}$ of the central velocity change, for the
      velocity interval of $-30 \la V_\mathrm{C} \la
      -10~\mathrm{km\,s}^{-1}$ this gradient increases at least to
      $1.5~\mathrm{km\,s}^{-1}$ per $1~\mathrm{km\,s}^{-1}$. Therefore,
      to restrict ourselves to the region of more or less constant mean
      line-widths, we accept for the following the velocity range
      appropriate for the higher latitude data and indicated in
      Fig.~\ref{Fig04}a by solid vertical lines. After such selection
      we separated in each class of profiles the Gaussians, central
      velocities of which fall into this range of $-9 \le V_\mathrm{C}
      \le 4~\mathrm{km\,s}^{-1}$, and constructed for every class the
      distribution of the widths of these components.

      \begin{figure}
         \resizebox{\hsize}{!}{\includegraphics{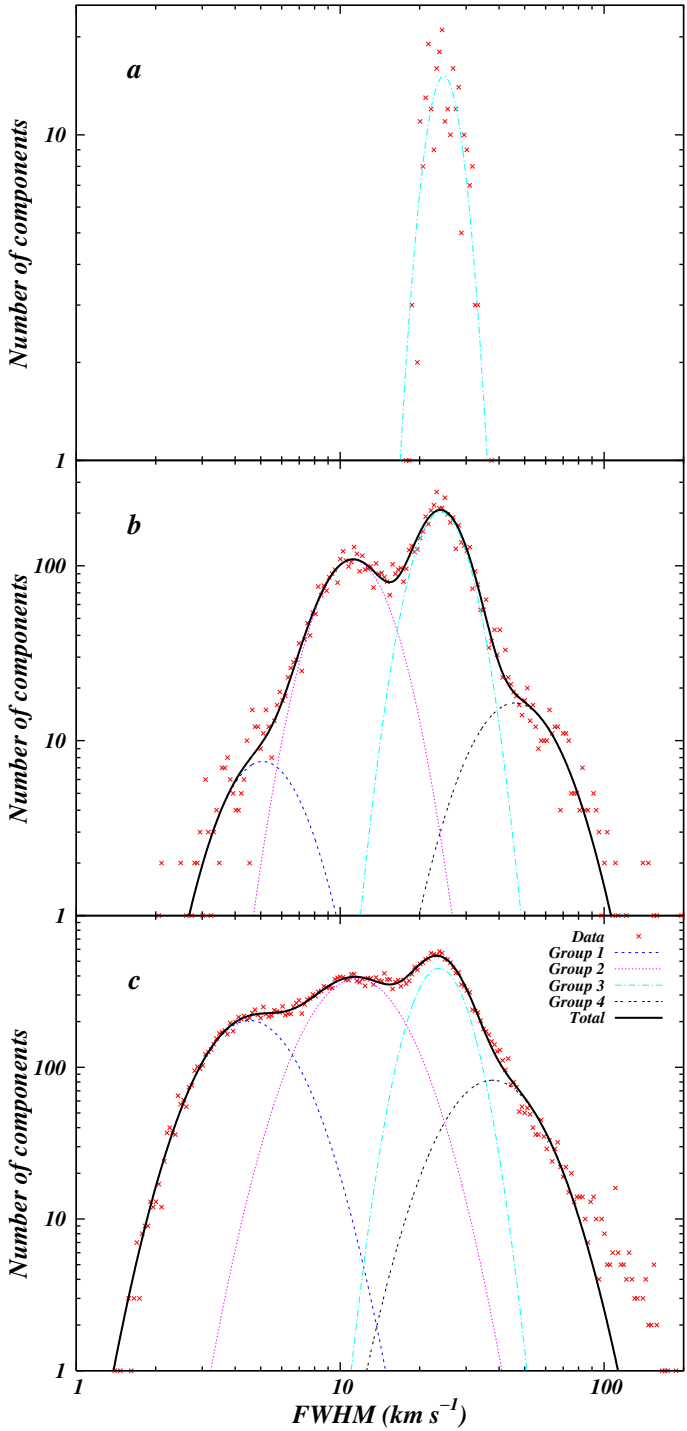}}
         \caption{The distribution of the widths of Gaussians for the
            profiles decomposed to one (upper panel), two (middle
            panel) or three (lower panel) components.}
         \label{Fig06}
      \end{figure}
      It turned out that for profiles decomposed into one, two or three
      accepted Gaussians, the concentration of the widths around some
      distinct mean values was rather well defined without any
      additional selection criteria (Fig.~\ref{Fig06}). To estimate the
      mean widths characteristic of these different line-width groups,
      we once again modelled the obtained distributions as sums of
      lognormal distributions. As in Paper~II we saw that the
      discrimination between the probably real and spurious Gaussians
      becomes more difficult at larger widths, we restricted the
      modelling process to the line-widths $\mathrm{FWHM} <
      100~\mathrm{km\,s}^{-1}$. Approximately above this width limit
      the selection criterion given by Eq. (\ref{Eq4}) becomes
      inoperative and at the greater widths no Gaussians are excluded
      by it. As a result, in these regions the proportion of the
      spurious components among the Gaussians, accepted for our
      analysis, may increase. This increase may be visible in the
      $\mathrm{FWHM} > 100~\mathrm{km\,s}^{-1}$ region of
      Fig.~\ref{Fig06}c as a group of points above the fitted model
      line.

      As can be seen from Fig.~\ref{Fig06}a, when the profile is
      successfully decomposed by only one low velocity Gaussian, such
      components belong exclusively to the line-width group~3, as
      defined by Fig.~\ref{Fig06}c. When the profile is decomposed by
      two Gaussians, the widths of corresponding components have
      already a much wider spread (Fig.~\ref{Fig06}b). Most of them
      still fall to group~3, but also to group~2, first discussed in
      connection with Fig.~\ref{Fig04}a, is well represented. Moreover,
      this is the case where group~2 is most clearly visible as a
      separate concentration of the line-widths. Beside these two
      dominating groups two weaker ones can be seen. On the left-hand
      side of the distribution, signs of group~1, which was the main
      group in Fig.~\ref{Fig04}b, become visible, but on the right-hand
      side the presence of a new group seems to be rather obvious.

      In Fig.~\ref{Fig06}c the role of group~1 has considerably
      increased and it has become a well-established component of the
      distribution. The new fourth group still remains rather
      ill-defined. Moreover, as with more complex profiles the role of
      extremely wide Gaussians increases, the presence of group~4
      becomes even somewhat more questionable than it may seem on the
      basis of Fig.~\ref{Fig06}b. Therefore, we add this new group to
      our list of the possible frequency enhancements in the
      distribution of the Gaussian widths, but we must remember that
      this group is established considerably less reliably than the
      other three.

      On the basis of the profiles decomposed into one, two or three
      low velocity Gaussians, the estimates for the mean line-widths of
      the four groups of the components described above, are $4.9 \pm
      0.1$, $12.0 \pm 0.3$, $24.4 \pm 0.2$, and $43 \pm
      5~\mathrm{km\,s}^{-1}$. Unlike the previous estimates, here the
      errors are not given on the basis of the formal errors of the
      parameter fitting, but a procedure somewhat similar to bootstrap
      was used. We estimated the mean values and their formal errors
      separately for profiles with one, two or three Gaussians and also
      for any combination of such classes (the profiles with one and
      two or one and three Gaussians together and so on), calculated
      the weighted (on the basis of formal fitting errors) means of the
      results and accepted as an error the standard deviation of the
      obtained estimates from these means. We expect that such
      estimates include more realistically the uncertainties of the
      obtained values than the pure formal errors of the fit of the
      model to a fixed dataset.

      For more complex profiles (four and more accepted Gaussians per
      profile) the components of the first group (the narrowest ones)
      become more and more dominating and this makes the recognition of
      the second group harder and harder (this is well demonstrated
      also by Fig.~\ref{Fig04}, where the second group is invisible in
      the lower panel, which represents the complex profiles near the
      galactic plane). The third group remains relatively well
      established, but the identification of the fourth group becomes
      increasingly complex and dependent on the accepted upper width
      limit, as in such profiles the frequency of extremely wide
      components increases. However, these changes are weaker for
      weaker profiles. At the same time, there is no clear-cut line
      between weak and strong profiles. Therefore, we decided to
      attempt to determine the width distributions of the Gaussians for
      different profile strength limits and we used in the following
      two different ways to define the profile strength. In the first
      case, we used as an indicator of the profile strength the peak
      brightness temperature of the profile (as estimated from the
      smooth Gaussian representation), but in the second case, we used
      the mean channel value (calculated from the original observed
      profile) over the velocity range used for decomposition.

      \begin{figure*}
         \sidecaption
         \includegraphics[width=12cm]{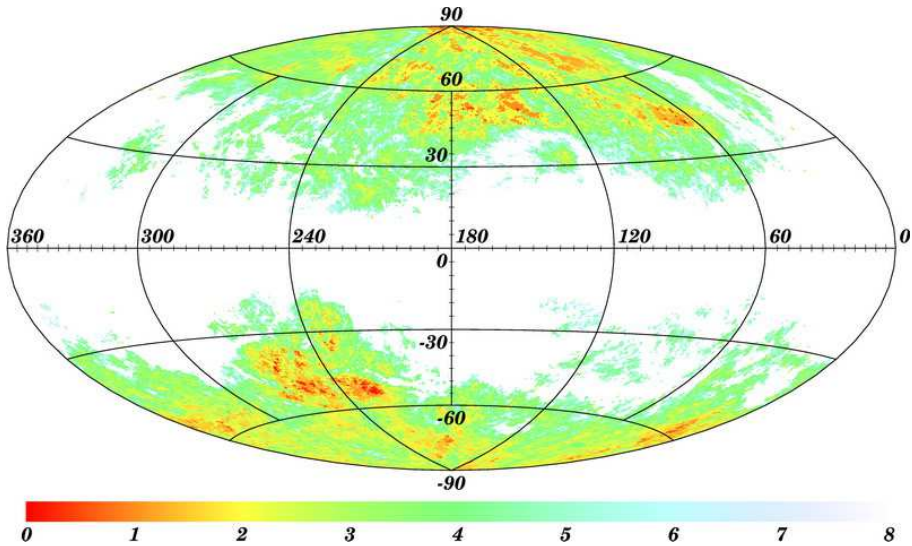}
         \caption{The sky positions of the 79\,648 profiles used for
            the final line-width estimates. The positions of the
            simplest profiles are marked by the reddest dots. The
            colour scale represents in the first order the number of
            Gaussians per profile. Among the profiles with an equal
            number of Gaussians those with lower peak brightness
            temperature are plotted with redder colour.}
         \label{Fig08}
      \end{figure*}
      For both profile strength definitions we repeated the estimates
      over a wide range of strength upper limits, but accepted for
      final results only the cases, where the means for all lognormal
      distributions were estimated with the formal errors below 50\%
      (or 25\% in a more restrictive case). In the case of the strength
      definition through the peak brightness temperature, the profile
      strength upper limit was increased in steps of $0.1~\mathrm{K}$
      from the zero until there were no more acceptable results during
      the last 10 steps. Only for the profiles decomposed into three
      Gaussians even the inclusion of the strongest lines with the
      highest peaks extending up to $44.3~\mathrm{K}$ permitted
      determination of the parameters of different line-width groups. 
      In all other cases no profiles with $T_\mathrm{b,max} >
      15.8~\mathrm{K}$ could be used and 89.7\% of the profiles used
      for the final estimates had $T_\mathrm{b,max} \le
      12~\mathrm{K}$.

      When using the peak brightness temperature as the profile
      strength indicator, the obtained mean line-widths were $4.3 \pm
      0.6$, $12.0 \pm 0.5$, $24.1 \pm 0.7$, and $42 \pm
      5~\mathrm{km\,s}^{-1}$ for the 50\% maximal error limit and $4.6
      \pm 0.5$, $12.1 \pm 0.4$, $24.1 \pm 0.6$, and $41 \pm
      4~\mathrm{km\,s}^{-1}$ for 25\%. Then the same procedure, but
      increasing the profile strength upper limit in steps of
      $0.001~\mathrm{K}$ was also repeated for the mean channel value
      as a profile strength indicator. In this case, the results were
      $3.9 \pm 0.6$, $11.7 \pm 0.4$, $24.1 \pm 0.6$, and $43 \pm
      5~\mathrm{km\,s}^{-1}$ for the 50\% limit and $4.3 \pm 0.7$,
      $11.8 \pm 0.4$, $24.1 \pm 0.7$, and $42 \pm
      4~\mathrm{km\,s}^{-1}$ for 25\%.

      As we can see, the results are in general agreement within their
      error estimates, which consider the dispersion of the results for
      a particular selection process, but do not take into account the
      uncertainties related to the choice of the peak brightness or
      mean channel value as the profile strength indicator or the usage
      of the 25\% or the 50\% precision limit. As the results seem to
      depend more on the way of estimating the profile strength than on
      the limit of formal errors of the parameter determination, we
      accept as the final result the average over different profile
      strength estimates for the 50\% error limit. After correcting the
      mean line-widths also for the width of the correlator channels in
      the LAB the corresponding values are: $3.9 \pm 0.6$, $11.8 \pm
      0.5$, $24.1 \pm 0.6$, and $42 \pm 5~\mathrm{km\,s}^{-1}$.
      Altogether these estimates are based on the decomposition of
      79\,648 selected profiles. This set of profiles represents 42.0\%
      of all observed sky positions in the LAB. The sky distribution of
      the corresponding profiles is given in Fig.~\ref{Fig08}.

      As an additional check, we repeated the procedure described in
      this section also for the symmetrical velocity limit
      ($|V_\mathrm{C}| \le 8~\mathrm{km\,s}^{-1}$ ) of the analysed
      profiles. As a profile strength indicator, we used the peak
      brightness and accepted the fits where the means for all
      lognormal distributions were estimated with the formal errors
      below 50\%. In this case, we obtained the following values: $4.3
      \pm 0.5$, $12.0 \pm 0.5$, $24.0 \pm 0.7$, and $43 \pm
      5~\mathrm{km\,s}^{-1}$, the results still in good agreement with
      the earlier ones.

      As we can see from Fig.~\ref{Fig08}, when selected on the basis
      of the number of Gaussians per profile and the strength of the
      emission line, the profiles used do not follow any well-defined
      latitude limit. Nevertheless, the mean values obtained for
      different line-width groups are still in good agreement with the
      estimates from the latitude limited samples obtained in Sec.~3.
      (we remind that the error estimates in these two cases are based
      on different considerations). This result, when combined also
      with Fig.~\ref{Fig04}, indicates that when there are clear
      differences in the overall line-width distribution of the most
      simple and most complex profiles, the transition from one to
      another is relatively smooth, without a well-defined boundary
      between them. Moreover, as differently selected samples give
      rather similar values for the mean line-widths of the groups, it
      seems that the width distributions of the profiles of different
      complexity differ from each other mainly by the degree of the
      presence of different line-width groups and not so much by the
      mean widths of these groups.

   \section{The comparison with earlier results}

      In the previous section we demonstrated that in the distribution
      of the widths of the Gaussians representing the simplest 21-cm
      line profiles of the galactic \ion{H}{i}, we can distinguish at
      least three, but probably even four distinct groupings of the
      line-widths. Earlier, similar results have been reported by
      Verschuur and co-authors, who have published several papers
      (\cite{Ver89}, \cite{Ver94}, \cite{Ver99} hereafter VP,
      \cite{Ver02}, \cite{Ver04} hereafter V4) where they argue that
      neutral hydrogen emission profiles produced by gas in the local
      interstellar medium are characterised by three, or probably four,
      line-width regimes where dominant and pervasive features have
      widths of the order $5.2 \pm 0.4$ (VP regime~3), $12.8 \pm 0.4$
      (VP regime~2), $31.0 \pm 0.5$ (VP regime~1b), and $50.1 \pm
      0.6~\mathrm{km\,s}^{-1}$ (VP regime~1a). They also
      note that these line-width regimes show a striking resemblance to
      a set of velocity regimes described by a plasma physical
      mechanism called the critical ionization phenomenon.

      As we can see, our results are qualitatively identical. We both
      clearly distinguish three different groupings or regimes of the
      \ion{H}{i} line-widths and admit that there may be also the
      fourth one with a larger mean line-width than the three main
      groups. However, we must recognize that there are considerable
      differences in the numerical values of the mean line-widths
      obtained by us and by Verschuur (as given in VP). On average, our
      results are about 17\% smaller than those given in VP. The
      differences are the largest (approximately 22\%) for groups~1 and
      3 (VP regimes~3 and 1b) and the smallest (12\%) for groups~2 and
      4 (VP regimes~2 and 1a). At the same time, there are also
      considerable differences in our approaches to data analysis. Our
      studies are based on somewhat different initial data (the LAB
      instead of the published version of the LDS), we have used
      different Gaussian decomposition algorithms and selection
      criteria for choosing the components for final analysis. Our
      approaches to estimating the mean line-widths of the groups of
      the components are also different. Therefore, it is interesting
      to check how all these differences may have influenced the
      results obtained. As in most cases in his papers Verschuur has
      not given the results for his regime~1a, we follow the same
      practice in this section.

      In Gaussian decomposition Verschuur and co-authors have clearly
      preferred the human assisted approach, as they assume that
      Gaussian decomposition ``is not something that can be left to an
      idealized computer program'' (VP). We have relied on the computer
      from the beginning (Paper~I). The probability that the
      differences in decomposition algorithms may be the primary source
      of the discrepancies in the mean line-widths of our line-width
      groups and Verschuur's regimes could be reduced if we could
      demonstrate that our programme gives for identical initial data
      practically the same results as obtained by Verschuur. This can
      be most easily done for the ``random directions chosen from the
      Leiden/Dwingeloo data files'' for which VP give in their Table~2
      the list of sky positions from where the profiles of the sample
      have been chosen.

      We analysed the Gaussians corresponding to these profiles of the
      LDS. In the width range of $0.5 < \mathrm{FWHM} <
      60~\mathrm{km\,s}^{-1}$, used by VP (see Fig.~4b in VP), we found
      three concentrations of the line-widths around mean values of
      $\mathrm{FWHM} = 5.4 \pm 0.7$, $13.0 \pm 0.5$ and $29.7 \pm
      6.8~\mathrm{km\,s}^{-1}$. In their Table~7, VP gave for the same
      sample the values equal to $5.2 \pm 1.3$, $12.8 \pm 1.2$ and
      $32.8 \pm 1.8~\mathrm{km\,s}^{-1}$. As can be seen, for exactly
      the same initial data our results coincide within errors (here we
      use once again the formal errors of the fit).

      The general selection criteria used by Verschuur are best
      described in the second chapter of V4. To test these criteria and
      to approximate the V4 results, we decomposed 206\,671 profiles of
      the LDS (as published by Hartmann \& Burton \cite{Har97}) into
      1\,644\,665 Gaussians and applied the following selection
      criteria defined in accordance with our understanding of the
      criteria used in V4:
      \begin{enumerate}
         \item Profiles along the lines of constant Galactic longitude
            at $10\degr$ intervals, starting with $l = 0\degr$ and
            every $1\degr$ in latitude were used.
         \item The latitude range was from $|b| = 80\degr$ (inc.) down
            toward the Galactic plane until:
         \begin{enumerate}
            \item the latitudes $|b| < 40\degr$ were reached, or
            \item the unsmoothed profile was found in which the
               brightness temperature in at least one channel in the
               velocity range of $-460 < V_\mathrm{C} <
               396~\mathrm{km\,s}^{-1}$ was $10~\mathrm{K}$ or greater.
         \end{enumerate}
         \item If the smooth profile, as reconstructed from the
            Gaussian decomposition, contained in the velocity interval
            of $-60 \le V_\mathrm{C} \le 40~\mathrm{km\,s}^{-1}$
            multiple peaks with separation less than
            $20~\mathrm{km\,s}^{-1}$ and the depth of the minimum
            between them less than 10\% of the height of the lower
            peak, the Gaussians corresponding to this profile were not
            used in the analysis.
         \item Any Gaussian with a peak brightness temperature of $<
            0.15~\mathrm{K}$ was removed from further consideration.
         \item Only the components with the central velocity
            $V_\mathrm{C}$ in the range of $-31 \le V_\mathrm{C} \le
            18~\mathrm{km\,s}^{-1}$ were used for the study and all
            profiles requiring more than four such Gaussians were
            removed from the analysis.
      \end{enumerate}

      Following V4, we analysed the results separately for positive and
      negative latitude data. For positive latitudes we obtained the
      mean values $\mathrm{FWHM} = 4.5 \pm 0.5$, $14.8 \pm 1.7$, $30.4
      \pm 1.9~\mathrm{km\,s}^{-1}$ and for negative-latitudes we found
      $\mathrm{FWHM} = 3.9 \pm 0.5$, $11.5 \pm 1.9$, $23.2 \pm
      0.4~\mathrm{km\,s}^{-1}$. We can see that on average these
      numbers are slightly larger than the corresponding results based
      on our selection criteria and the LAB data. Most likely this is
      due to the wider velocity range used by V4 (indicated in
      Fig.~\ref{Fig04}a by two dashed vertical lines). However, in the
      first order, these numbers must be compared with those given in
      Table~7 of V4: $\mathrm{FWHM} = 5.8 \pm 2.4$, $13.7 \pm 2.3$,
      $35.1 \pm 4.0~\mathrm{km\,s}^{-1}$ for positive and
      $\mathrm{FWHM} = 5.2 \pm 1.9$, $14.5 \pm 2.4$, $33.9 \pm
      3.6~\mathrm{km\,s}^{-1}$ for negative latitude data, respectively
      (here our and V4 error estimates are most likely not directly
      comparable). With the exception of group~2, our values are still
      smaller than those by V4.

      So far we have estimated the mean line-widths of the different
      line-width groups through the approximation of the distribution
      by the sum of lognormal functions. At the same time, the mean
      values given in Table~7 of V4 are calculated as averages over the
      predefined ranges of widths. If we apply the same approach and
      the same line-width ranges to our results described in this
      chapter, our mean line-widths are for positive-latitude data
      $\mathrm{FWHM} = 5.7 \pm 2.2$, $14.3 \pm 2.8$, $34.4 \pm
      4.7~\mathrm{km\,s}^{-1}$ and for negative-latitude data
      $\mathrm{FWHM} = 5.6 \pm 2.2$, $14.2 \pm 3.0$, $34.4 \pm
      4.9~\mathrm{km\,s}^{-1}$ (here the standard deviations of the
      line-widths around the means in predefined width intervals are
      given as error estimates as seems to have been done also by V4).
      These numbers are in good agreement with the V4 results as cited
      above. However, in this case it remains somewhat questionable
      how justified the averaging ranges used in V4 are. For example,
      why are in Table~7 of V4 the Gaussians with widths of $20 <
      \mathrm{FWHM} < 28~\mathrm{km\,s}^{-1}$ not included into any
      line-width regime?

   \section{Discussion}

      In V4 the obtained mean line-widths are used to estimate the
      kinetic temperatures of the gas. Verschuur found that the
      temperatures corresponding to the line-width regimes~1b and 3 are
      $24\,000~\mathrm{K}$ and $750~\mathrm{K}$, respectively. These
      temperatures were considered as physically unreal. On the basis
      of the coincidence of the obtained mean line-widths with the
      critical ionization velocities (CIV) for some atomic species
      ($6~\mathrm{km\,s}^{-1}$ for Na and Ca,
      $13.5~\mathrm{km\,s}^{-1}$ for C, N, and O,
      $34~\mathrm{km\,s}^{-1}$ for He and $51~\mathrm{km\,s}^{-1}$ for
      H, respectively) a model with collisionless gas, in which the
      origin of the profile broadening represents the influence of the
      critical ionization velocity effect, is presented. This
      interpretation has not found wide acceptance in astronomical
      community and our results make such explanation even more
      unlikely as the agreement between the selected CIVs and the mean
      line-widths seems to be poorer than stated by VP.

      However, some decades ago, Clark (\cite{Cla65}) proposed a
      two-component model for interstellar \ion{H}{i} where cold
      absorbing clouds are surrounded by hot, ubiquitous intercloud
      gas. Theoretical studies (e.g. Field et al. \cite{Fie69}) also
      demonstrated that the local \ion{H}{i} could be considered as a
      two-phase medium, where much of the gas is either a warm neutral
      medium (WNM) with $T \sim 8000~\mathrm{K}$ or a cold neutral
      medium (CNM) with $T \sim 50~\mathrm{K}$ (Kulkarni \& Heiles
      \cite{Kul87}; Dickey \& Lockman \cite{Dic90}; Wolfire et al.
      \cite{Wol95}; Ferri\`ere \cite{Fer01}). At present this is one of
      the most widely accepted models for the local \ion{H}{i}.
      Therefore, we try to check how our four line-width groups may be
      related to this model.

      Mebold (\cite{Meb72}) has estimated that the emission line
      components, corresponding to the WNM, have a Gaussian dispersion
      of about $9.7~\mathrm{km\,s}^{-1}$ or $\mathrm{FWHM} \approx
      23~\mathrm{km\,s}^{-1}$. This width is rather close to the mean
      width of our third line-width group and therefore we may expect
      that this line-width group may be related to the presence of the
      WNM in the ISM. With other line-width groups the situation seems
      to be somewhat more complicated.

      Verschuur (\cite{Ver74}) and Schwarz \& van Woerden
      (\cite{Sch74}) have demonstrated that the histogram of the CNM
      \ion{H}{i} emission line widths peak around $\sigma_V =
      2.2~\mathrm{km\,s}^{-1}$, which corresponds to $\mathrm{FWHM}
      \approx 5.2~\mathrm{km\,s}^{-1}$ -- a value somewhere between our
      first and second line-width groups and directly not comparable to
      any line-width group. However, here we must notice that their
      results have been obtained using all narrow emission line
      components presented in the observed profiles. At the same time,
      as discussed in the Introduction, for stronger emission
      components the saturation effects become important and this makes
      the line shape nongaussian and such lines seem to be broader than
      unsaturated profiles.

      In our analysis we have used only relatively weak lines with
      nearly 90\% of the used profiles having peak brightness less than
      $12~\mathrm{K}$. By combining the equation of radiative transfer
      with the approximate correlation between the \ion{H}{i} spin
      temperature $T_\mathrm{S}$ and optical depth $\tau$ (e.g. Payne
      et al. \cite{Pay83}), we may estimate that for most of our
      profiles $\tau < 0.1$, and therefore, we may use the
      approximation for optically thin gas $T_\mathrm{b}(\nu) =
      T_\mathrm{S} \tau(\nu)$. For $\tau \le 0.1$ this approximation
      introduces up to 5\% error in the peak brightness temperature of
      the emission profile, but practically does not change the
      line-width and consequently we must compare our line-widths not
      with those of much stronger emission lines cited above, but with
      the line-widths of the absorption lines of the CNM.

      Crovisier (\cite{Cro81}) has reported that the distribution of
      the velocity dispersion inside the CNM clouds peaks around
      $1~\mathrm{km\,s}^{-1}$, but the mean value for the distribution
      is about $1.7~\mathrm{km\,s}^{-1}$. The latter corresponds to
      $\mathrm{FWHM} \approx 4.0~\mathrm{km\,s}^{-1}$, which coincides
      within errors with the mean line-width of our first line-width
      group. This makes it rather plausible that the existence of both
      line-width groups~1 and 3 is a reflection of the presence of two
      thermal phases of neutral gas in the local interstellar medium.

      Theoretical CNM and WNM temperatures are derived by calculating
      the equilibrium temperature as a function of thermal pressure. As
      pointed out above, there exist two stable ranges of equilibrium,
      separated by a region of unstable temperatures. If we interpret
      the line-width of group~2, $11.8~\mathrm{km\,s}^{-1}$, as
      thermal, the corresponding kinetic temperature would be about
      $3000~\mathrm{K}$, which lies just in the region of unstable
      temperatures. At the same time, the presence of some gas with
      such temperatures seems to have been a long-standing problem.
      Cox (\cite{Cox05}) has recently given a review of the three-phase
      model of the interstellar medium, but there the additional third
      phase (warm \ion{H}{i}a) has a temperature of $5000~\mathrm{K}$.
      Heiles (\cite{Hei01}) and Heiles \& Troland (\cite{Hei03}) have
      reported that at least 48\% of the WNM lies in the thermally
      unstable region of $500-5000~\mathrm{K}$. Similar results were
      obtained by Mebold et al. (\cite{Meb82}). On the basis of the
      results by Dickey et al. (\cite{Dic77,Dic78,Dic79}), Crovisier
      (\cite{Cro81}) discussed the presence of cold clouds, lukewarm
      clouds and not strongly absorbing material. Therefore, with some
      probability our line-width group~2 may also represent thermally
      unstable neutral gas and our (also Verschuur's) results
      (Fig.~\ref{Fig04}a and especially Fig.~\ref{Fig06}b) demonstrate
      that this is not just a tail of the CNM or the WNM temperature
      distribution, but a component in its own rights.

      Besides first three rather obvious line-width groups there is
      also the line-width group~4 with the mean line-width around
      $42~\mathrm{km\,s}^{-1}$. As mentioned above, the real existence
      of this group is much more questionable, as it may be seriously
      affected by baseline and stray radiation corrections in the
      survey (see Paper~2). Therefore, we make only some brief remarks
      about this group. Kalberla et al. (\cite{Kal98}) have argued in
      favour of the existence of some neutral gas in the galactic halo.
      The mean velocity dispersion of this gas is $\sigma = 60 \pm
      3~\mathrm{km\,s}^{-1}$ at the north galactic pole and it
      increases at lower latitudes. The corresponding \ion{H}{i}
      line-widths must be more than 3 times higher than the mean for
      our line-width group~4. Line-widths (with $\sigma = 21 \pm
      4~\mathrm{km\,s}^{-1}$) much more similar to our group~4 are
      reported from the measurements of absorption lines by Dwarakanath
      (\cite{Dwa04}) and Mohan et al. (\cite{Moh4a,Moh4b}).

      It is also interesting to estimate the mass fractions of the
      \ion{H}{i} gas, corresponding to the described four line-width
      groups. However, in our case these estimates are made rather
      questionable by at least two factors. First of all, the line
      components with highest peak brightness temperatures belong to
      the line-width group~1 and in the following groups the components
      are on average progressively weaker. As for our estimates we have
      selected only the simplest and the weakest profiles, we have
      biased the mass fraction estimates in favour of the groups with
      wider and weaker line components. Therefore, for the estimates
      of mass fractions, it is more natural to return to our latitude
      limited samples (similar to Fig.~\ref{Fig04}a), but even in this
      case the stronger lines near the galactic plane become optically
      thicker, which causes the underestimation of corresponding
      masses.

      The second problem is that when the mean line-widths of our
      line-width groups are well determined by log-normal fits, the
      standard deviations of the corresponding distributions are rather
      poorly constrained due to correlations between these parameters
      for different line-width groups. To obtain the fits for different
      latitude limits $|b| \ge b_\mathrm{lim}$, we had therefore to
      restrict the variability of these parameters and this probably
      introduces some systematic errors into the mass fraction
      estimates, specially for lower values of $b_\mathrm{lim}$.

      Nevertheless, we found that for $|b| \ge 40\degr$ the velocity
      interval of $-9 < V_\mathrm{C} < 4~\mathrm{km\,s}^{-1}$ contains
      more than 51\% of all \ion{H}{i}. As the line-width groups found
      in this velocity interval seem to continue even beyond the
      accepted velocity limits (with the exception of group~2, which is
      visible only in the vicinity of $V_\mathrm{C} =
      0~\mathrm{km\,s}^{-1}$, they just become smoothly wider at higher
      velocities), we may conclude that the described line-width
      concentrations are characteristic of most of the high-latitude
      \ion{H}{i}.

      The distribution of the \ion{H}{i} between the obtained
      line-width groups depends on the latitude limit used. Only the
      relative content of group~4 is about 4\% for all samples $|b| \ge
      b_\mathrm{lim}$ with different values of $b_\mathrm{lim}$. For
      $15\degr \ga b_\mathrm{lim} \ga 55\degr$ the relative fractions
      of other groups change rather slowly and smoothly. The share of
      groups~1 and 2 decreases from about 18\% and 45\% for $|b| \ge
      15\degr$ to about 11\% and 39\% at $|b| \ge 55\degr$,
      respectively. However, we must stress once more that the
      estimates for these groups are strongly correlated and especially
      for lower galactic latitudes the actual fraction of group~1 may
      be higher and that of group~2 lower than stated here. The content
      of group~3 increases for the same latitude range from about 34\%
      to 46\%. For larger values of $b_\mathrm{lim}$ the relative
      fraction of group~1 remains at about 11\%, but the share of
      group~2 quickly increases to about 47\% at $b_\mathrm{lim} =
      75\degr$ and that of group~3 drops to about 37\%. Therefore, for
      most of the relatively high-latitude gas we, like Heiles
      (\cite{Hei01}), may conclude that about half of the WNM is
      represented by line-width group~2 which temperatures lie in the
      thermally unstable regime. For $b_\mathrm{lim} < 15\degr$ our
      analysis is most likely not applicable at all and for
      $b_\mathrm{lim} \ge 75\degr$ the results become uncertain due to
      the small number of profiles.

   \section{Conclusions}

      We have demonstrated that for simple (Sec.~4.) and high-latitude
      (Sec.~3.) profiles the line-widths of the \ion{H}{i} emission
      profiles exhibit concentration to certain preferred line-width
      groups. With this we confirm and extend the earlier similar
      results obtained by Verschuur and co-authors (\cite{Ver89},
      \cite{Ver94}, \cite{Ver99}, \cite{Ver04}). We define the limits
      of the usable velocities ($-9 \le V_\mathrm{C} \le
      4~\mathrm{km\,s}^{-1}$) on the basis of the extent of the region,
      where the dependence of the line-widths on the component
      velocities is the weakest. We have not fixed any exact criteria
      for the simplicity of the profiles, but have tried instead many
      different combinations of the limits on the number of the
      Gaussians in the decomposition and on the profile strength as
      defined by their peak brightness or mean channel value. For the
      final results we have accepted only these combinations of the
      limits, which allow the determination of the mean line-widths of
      all width groups with the formal fitting errors less than 50\%.
      This condition was fulfilled for all profiles, decomposed into
      one, two or three components. For such profiles we demonstrated
      that it is possible to distinguish three or four groups of
      preferred line-widths. The existence of groups~2 and 3 is very
      clearly visible, when using only the profiles decomposable by one
      or two Gaussians (Figs.~\ref{Fig06}a and \ref{Fig06}b). Group~1
      can be better distinguished in profiles decomposed by three
      components (Fig.~\ref{Fig06}c). The reality of group~4 remains
      more questionable. A similar separation was possible also for
      weaker profiles decomposed into four, or in some cases even into
      up to 7 Gaussians. The mean line-widths of the groups found are
      $3.9 \pm 0.6$, $11.8 \pm 0.5$, $24.1 \pm 0.6$, and $42 \pm
      5~\mathrm{km\,s}^{-1}$.

      The mean line-widths obtained by us are somewhat smaller than
      those proposed earlier by Verschuur. The disagreement may be
      caused partly by different algorithms used to measure the mean
      line-widths of our groups and VP regimes. Another important point
      appears to be the usage by V4 of Gaussians from a considerably
      wider velocity range than it is acceptable to us. Due to the
      smaller mean line-widths of our line-width groups, also the
      interpretation of the existence of the corresponding groups
      through the collisionless gas and critical ionization velocities,
      as proposed by Verschuur, seems rather questionable to us. More
      likely groups~1 and 3 represent CNM and WNM components of the
      ISM. The existence of group~2, which may represent up to 40\% of
      all \ion{H}{i}, in the thermally unstable regime, is in conflict
      with usual two component equilibrium models of the ISM, but from
      our data (Figs.~\ref{Fig04}a and \ref{Fig06}b) the existence of
      such a group seems to be well established. The reality and the
      origin of the broad lines with the widths of about
      $42~\mathrm{km\,s}^{-1}$ is more obscure. These, however, contain
      only about 4\% of the total observed column densities.

   \begin{acknowledgements}
      We would like to thank W.~B.~Burton for providing the preliminary
      data from the LDS for programme testing prior the publication of
      the survey. A considerable part of the work on creating the
      decomposition programme was done during the stay of U.~Haud at
      the Radioastronomical Institute of Bonn University (now
      Argelander-Institut f\"ur Astronomie). The hospitality of the
      staff members of the Institute is greatly appreciated. We thank
      G.~L.~Verschuur from the University of Memphis, E.~Saar from
      Tartu Observatory and the anonymous referee for fruitful
      discussions and considerable help. The project was supported by
      the Estonian Science Foundation grant no. 6106.
   \end{acknowledgements}

\end{document}